\documentclass[a4paper]{jpconf}
\usepackage{citesort,amsmath,graphicx,iopams}

\pdfoutput=1

\newcommand{\be}{\begin{equation}}
\newcommand{\ee}{\end{equation}}
\newcommand{\bea}{\begin{eqnarray}}
\newcommand{\eea}{\end{eqnarray}}

\begin{document}
\title{Towards a quantitative understanding of high $p_T$ flow harmonics}

\author{Jorge Noronha}
\address{Instituto de F\'{\i}sica, Universidade de S\~{a}o Paulo, C.P. 66318,
05315-970 S\~{a}o Paulo, SP, Brazil}

\ead{noronha@if.usp.br}

\begin{abstract}
In this proceedings I briefly review the recent progress achieved on the calculation of $v_n$ at high $p_T$ via the coupling of a jet energy loss model with full event-by-event viscous hydrodynamics. It is shown that that this framework can simultaneously describe experimental data for $R_{AA}$, $v_2$, and $v_3$ at high $p_T$. High $p_T$ $v_2$ is found to be approximately linearly correlated with the soft $v_2$ on an event-by-event basis, which opens up a new way to correlate soft and hard observables in heavy ion collisions.
\end{abstract}

\section{Introduction}

The Quark-Gluon Plasma (QGP) formed in heavy ion collisions is the smallest (and the hottest) most perfect fluid ever made. Arguably, evidence for the formation of the QGP comes from three fronts: (i) the equation of state (EOS) computed using \emph{lattice QCD} \cite{Borsanyi:2013bia} shows that at temperatures $T> 200$ MeV (easily achievable in these collisions) quarks and gluons are not confined into hadrons, (ii) the large \emph{anisotropic flow} of low $p_T$ hadrons, computed using event-by-event viscous hydrodynamics simulations (for reviews see \cite{Heinz:2013th,Luzum:2013yya}), requires the formation of a nearly inviscid medium that is not consistent with purely hadronic expectations \cite{NoronhaHostler:2008ju,NoronhaHostler:2012ug,Denicol:2013nua}, (iii) the large suppression of high $p_T$ hadrons in AA collisions with respect to elementary pp collisions, a simple consequence of in medium \emph{jet quenching} \cite{Gyulassy:1990ye,Wang:1991xy}. The three items above are not disconnected. For instance, the EOS is used in the hydrodynamic modeling of the evolving QGP from which anisotropic flow coefficients are computed and this hydrodynamically expanding fluid serves as a background for the passage of jets.  

While the QGP equation of state at zero baryon chemical potential is under control \cite{Bazavov:2014pvz} and current event-by-event hydrodynamic simulations have achieved unprecedented levels of success \cite{Gale:2012rq} (and predictive power \cite{Noronha-Hostler:2015uye,Niemi:2015voa}), the mechanism involving both the soft and hard phenomena responsible for the generation of anisotropic flow at high $p_T$ remained somewhat elusive (see the discussion in \cite{Betz:2014cza}) and only now some of its features have become clearer. For instance, it was shown in \cite{Noronha-Hostler:2016eow,Betz:2016ayq} that realistic event-by-event hydrodynamic modeling plays an important role in solving the high $p_T$ $R_{AA} \otimes v_2$ puzzle (see \cite{Betz:2014cza} and refs.\ therein) in heavy ion collisions.

\section{Results from the event-by-event viscous hydrodynamics + jet energy loss model}

In this proceedings we use the v-USPhydro code \cite{Noronha-Hostler:2013gga,Noronha-Hostler:2014dqa,Noronha-Hostler:2015coa} to model the expanding QGP fluid produced in PbPb collisions at $\sqrt{s}=2.76$ TeV and solve the viscous hydrodynamic equations event-by-event. The details about the parameters of the hydrodynamic calculations can be found in \cite{Noronha-Hostler:2016eow}. In order to vary the size of the initial energy density eccentricities present in the initial conditions for hydrodynamics, we used MCGlauber and MCKLN initial conditions \cite{Drescher:2006ca} for the mid-central $20-30\%$ centrality class at the LHC. Our results for the low $p_T$ soft $v_2$ and $v_3$ provide a good description of the data \cite{Chatrchyan:2013kba} (the shear viscosity over entropy density ratio is $\eta/s=0.08$ in MCGlauber and $\eta/s=0.11$ in MCKLN), as one can see in Fig.\ 1 of \cite{Noronha-Hostler:2016eow}. Once the hydrodynamic background is fixed and the low $p_T$ azimuthal anisotropies are reproduced, one can use the spacetime profile of the hydrodynamic fields, computed event-by-event, in energy loss calculations.    

In this work the azimuthally averaged nuclear modification factor, $R_{AA}(p_T)$, and the high $p_T$ azimuthal anisotropies (defined via a Fourier expansion of $R_{AA}(p_T,\phi)$) are investigated using the BBMG jet-energy loss \cite{Betz:2011tu,Betz:2012qq,Betz:2014cza}. In this model, the parton energy loss per unit length, $dE/dL$, is modeled as $\frac{dE}{dL} = -\kappa\,
E^a(L) \,L^z\, T^c \,\zeta_q\, \Gamma_{\rm flow}$, where $\kappa$ is the jet-medium coupling for quarks and gluons \cite{Betz:2014cza}, $T$ is the local temperature field along the jet path in the medium (with $c=2+z-a$), $\zeta_q$ describes energy loss fluctuations \cite{Betz:2014cza}, and the flow factor $\Gamma_{\rm flow}$ takes into account the boost from the local rest frame of the fluid. The parameters of the energy loss rate correspond to the pQCD-case defined in \cite{Betz:2014cza} $(a=0,z=1,c=3,q=0)$, which gives a linear path length dependence for the energy loss, $dE/dL \sim L$. Effects from a quadratic path length dependence were investigated in \cite{Betz:2016ayq}. 

\begin{figure}[ht]
\centering
\includegraphics[width=0.4\textwidth]{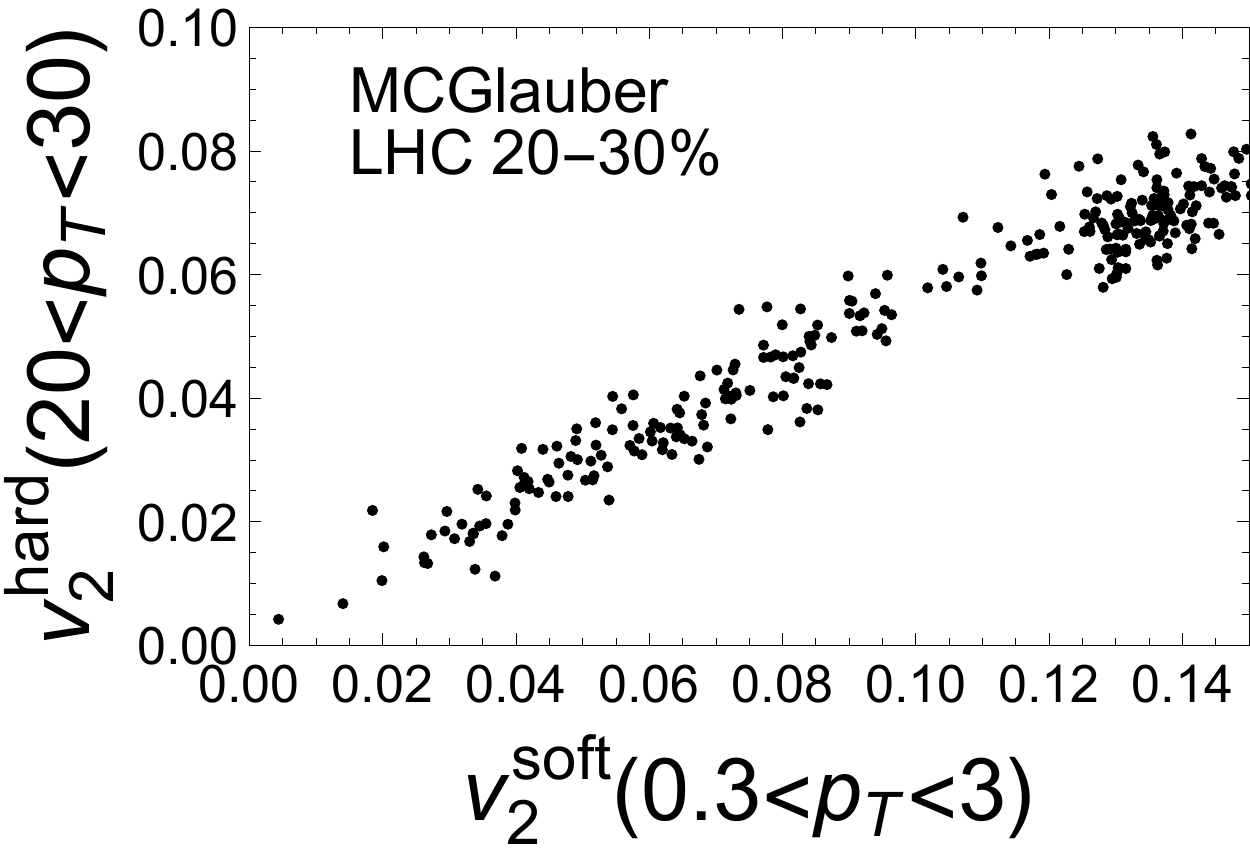}
\caption{Correlation between the soft and the hard elliptic flow $v_2$ event-by-event.}
\label{fig1}
\end{figure}

An important issue regarding the calculation of high $p_T$ azimuthal coefficients is that these coefficients are defined via a correlation between soft and hard particles over many events, which necessarily implies that the geometrical fluctuations present in the soft sector are carried over to the hard sector. In fact, while the initial state eccentricities drive the soft flow harmonics in hydrodynamics through pressure and flow gradients, azimuthal momentum anisotropies at high $p_T$ carry information about the initial state due to differences in the path length. Thus, one expects to find an approximate linear correlation between the soft and the hard elliptic flows event-by-event, since both are generated by the fluctuating initial spatial eccentricity $\varepsilon_2$. Such a linear correlation can be clearly seen in our model calculations displayed in Fig.\ \ref{fig1} where each point corresponds to a hydro event (which contains many jets). This behavior motivated the more detailed study involving 4-particle cumulants of high $p_T$ elliptic flow, which involves one hard particle correlated with 3 soft ones, performed in \cite{Betz:2016ayq}. 

\begin{figure*}[ht]
\centering
\includegraphics[width=1\textwidth]{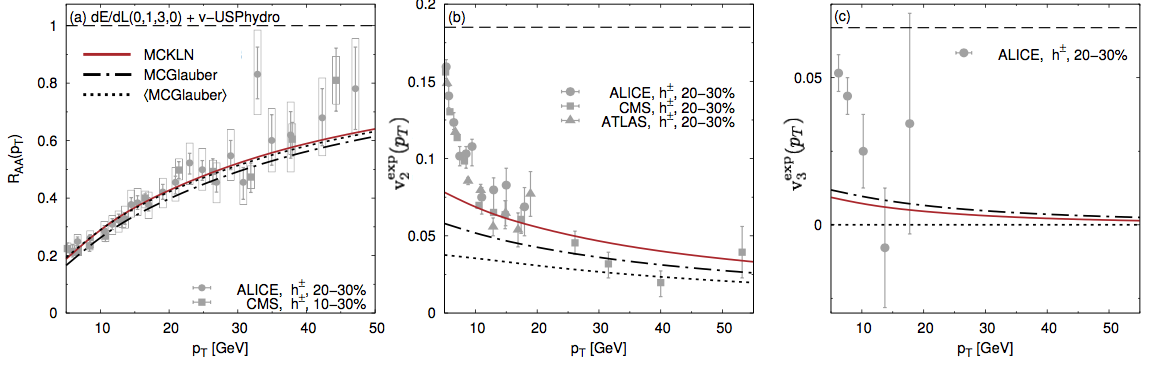}
\caption{(Color online) Comparison between our results for the 
$\pi^0$ $R_{AA}(p_T)$, (b) $v_2^{exp}(p_T)$, (c) $v_3^{exp}(p_T)$ in 
mid-central $\sqrt{s}=2.76$ TeV Pb+Pb collisions at the LHC to data \cite{ALICE_RAA,CMS_RAA,ALICE_v2_v3,CMS_v2,ATLAS_v2}.}
\label{fig2}
\end{figure*}

A comparison between our results for the $\pi^0$ $R_{AA}(p_T)$, (b) $v_2^{exp}(p_T)$, (c) $v_3^{exp}(p_T)$ in mid-central $\sqrt{s}=2.76$ TeV Pb+Pb collisions at the LHC and the experimental data \cite{ALICE_RAA,CMS_RAA,ALICE_v2_v3,CMS_v2,ATLAS_v2} can be found in Fig.\ \ref{fig2}. In these plots, $v_2^{exp}$ and $v_3^{exp}$ are computed via a soft-hard 2-particle correlator \cite{Luzum:2013yya,Betz:2016ayq}, which correctly takes into account the way the high $p_T$ measurement is performed. MCKLN initial conditions are solid red curves while the dotted-dashed black line stands for MCGlauber calculations. The black dotted line $\langle{\rm MCGlauber}\rangle$ corresponds to an event averaged smoothed initial Glauber geometry, shown here for comparison. One can see that the results computed using MCKLN initial conditions provide a good description of the data. In fact, given that the high $p_T$ flow harmonics are also determined by the initial eccentricities, initial conditions with an $\varepsilon_2$ larger than that found in MCGlauber, such as MCKLN (or IP-Glasma \cite{Gale:2012rq}), should provide a good description of $v_2$ at high $p_T$. On the other hand, the initial triangularity $\varepsilon_3$ is generally anti-correlated with $\varepsilon_2$ and this is why in Fig.\ \ref{fig2} $v_3$ in MCGlauber is a bit larger than that found using MCKLN initial conditions. 

\section{Outlook}

With the advent of realistic event-by-event viscous hydrodynamics + energy loss calculations \cite{Noronha-Hostler:2016eow,Betz:2016ayq}, it has become possible for the first time to simultaneously describe $R_{AA}$, $v_2$, and $v_3$ at high $p_T$ in heavy ion collisions. This validates the idea \cite{Gyulassy:2000gk} that jet energy loss determines the azimuthal anisotropy of the QGP at high $p_T$, improving our understanding about the dynamical features of the strongly coupled, deconfined matter formed in heavy ion collisions.  

Many obvious improvements in our current model are needed. On the hydro side, other initial conditions such as IP-Glasma \cite{Gale:2012rq} and Trento \cite{Moreland:2014oya} could be implemented and  one needs to investigate the effects of a $T$-dependent $\eta/s$ and bulk viscosity \cite{Ryu:2015vwa,Bernhard:2016tnd} as well. In this regard, we remark that the effect of higher order transport coefficients \cite{Finazzo:2014cna} in the dynamical evolution of the non-conformal strongly coupled QGP remains largely unexplored. Furthermore, full 3+1 hydrodynamic evolution would be needed to compute the rapidity dependence of high $p_T$ flow harmonics, in contrast to the boost-invariant scenario implemented here.    

Regarding the energy loss model, even though our current implementation is extremely simplistic, it does seem to possess the necessary features to describe high $p_T$ flow harmonics. With the new LHC data at $\sqrt{s}=5.02$ TeV one may be able to distinguish energy loss models with linear path length dependence from models where $dE/dL \sim L^2$, as shown in \cite{Betz:2016ayq}. A necessary next step involves using more realistic energy loss models which contain the expected weak coupling QCD features  as well non-perturbative effects from various sources (see \cite{Xu:2015bbz}). It would be interesting to see how realistic event-by-event viscous hydrodynamic modeling affects the flow harmonics in the heavy flavor sector (a preliminary study has been done in \cite{Prado:2016xbq}). 

By properly taking into account the effect of initial state fluctuations in the hydrodynamic medium, new observables \cite{Betz:2016ayq} involving the correlation between soft and hard flow harmonics can now be investigated, in contrast to all the previous calculations that employed unrealistic, event-averaged hydrodynamic backgrounds. It would be interesting to see if the overall distribution of high $p_T$ flow harmonics is similar to the one obtained at low $p_T$. At the moment, only calculations of the 4-particle cumulant $v_2\{4\}$ at high $p_T$ \cite{Betz:2016ayq} have been performed and higher order cumulants would be needed to assess the information contained in the event-by-event distributions. Better theoretical control of the fluctuations of flow harmonics at high $p_T$ can be useful to distinguish between different energy loss models. 

The type of analysis performed here must also be done in different collisions systems and different energies (see \cite{Betz:2016ayq} for the case of PbPb at $\sqrt{s}=5.02$ TeV). A challenging feat would be to perform realistic jet energy loss + event-by-event hydrodynamic calculations that can be used to simultaneously investigate the soft and the hard flow harmonics in small systems, such as pA collisions. Finally, concerning the complete understanding of flow harmonics in heavy ion collisions, one may now say that we have a good (quantitative) understanding of the underlying mechanisms responsible for the observed azimuthal anisotropies at low $p_T < 3$ GeV and high $p_T>10$ GeV. The hardest problem of quantitatively describing the non-monotonic behavior of flow harmonics at intermediate $p_T$, which I in jest have called the ``uncanny valley", requires a novel self-consistent way to couple jets with hydrodynamics on an event-by-event basis that goes way beyond the (modest) attempt pursued in \cite{Andrade:2014swa}. The solution to this problem remains, to the best of my knowledge, unknown. Perhaps some of the brave young minds that have contributed to make Hot Quarks 2016 a wonderful experience will lead the way towards solving this problem. 

\section*{Acknowledgements} J.N. thanks the University of Houston for its hospitality and FAPESP and CNPq for support.

\section*{References}
\bibliography{mybib}{}

\end{document}